\documentclass[useAMS,usenatbib]{mn2e}
\usepackage{psfig}
 \usepackage{txfonts}
 \usepackage{graphicx}
\usepackage{multirow}
\usepackage{lscape}
\usepackage{rotating}
 \title[A broad O VIII line in UCXB 4U 1543-624]
{Discovery of a broad O VIII Ly$\alpha$ line in the ultra-compact X-ray binary 4U 1543-624} 

\author [Madej et al.]
{O.K. Madej,$^{1,2}\thanks{E-mail: O.Madej@sron.nl}$ P.G. Jonker,$^{2,3,4}$ \\
\normalsize{$^{1}$Astronomical Institute, Utrecht University, PO Box 80000, 3508 TA Utrecht, The Netherlands}\\
\normalsize{$^{2}$SRON Netherlands Institute for Space Research, Sorbonnelaan 2, 3584 CA Utrecht,The Netherlands}\\
\normalsize{$^{3}$Harvard-Smithsonian Center for Astrophysics, 60 Garden Street, Cambridge, MA 02138, USA}\\
\normalsize{$^{4}$Radboud University Nijmegen, P.O. Box 9102, 6500 HC Nijmegen, The Netherlands} }

\begin{document}

\date{}

\maketitle
\label{firstpage}
\def\apjl{ApJ}
\def\aj{AJ}
\def\apj{ApJ}
\def\pasp{PASP}
\def\spie{SPIE}
\def\apjs{ApJS}
\def\araa{ARAA}
\def\aap{A\&A}
\def\nat{Nature}
\def\mnras{MNRAS}
\def\prd{Phys.Rev.D}
\begin{abstract}
We report the discovery of a broad emission feature at $\sim0.7$ keV in the spectra of the ultra-compact X-ray binary 4U 1543-624, obtained with the high-resolution spectrographs of the {\it XMM-Newton} and {\it Chandra} satellites. We confirm the presence of the feature in the broad band MOS2 spectrum of the source. As suggested before in the literature, the donor star in this source is a CO or ONe white dwarf, which transfers oxygen-rich material to the accretor, conceivably a neutron star. The X-rays reprocessed in this oxygen-rich accretion disc could give a reflection spectrum with O VIII Ly$\alpha$ as the most prominent emission line. Apart from the feature at $\sim 0.7$ keV we confirm the possible presence of a weak emission feature at $\sim 6.6$ keV, which was reported in the literature for this data set. We interpret the feature at $\sim 0.7$ keV and $\sim 6.6$ keV as O VIII Ly$\alpha$ and Fe K $\alpha$ emission respectively, caused by X-rays reflected off the accretion disc in the strong gravitational field close to the accretor. 
\end{abstract}

\vspace{-7mm}
\section{Introduction}
The ultra-compact X-ray binaries (UCXB)  are a subclass of low mass X-ray binaries (LMXB), which have very short orbital periods, typically less than 80 min. In this type of systems the hydrogen-poor donor star transfers material to a black hole or a neutron star. 
The UCXB candidate 4U 1543-624 has an orbital period of 18.2 min \citep{wang}. The source is placed near the Galactic plane ($l=322^{\circ}, b=-6^{\circ}$) at a distance of $\sim7$ kpc. It is a low inclination system ($i< 70^{\circ}$), since it lacks eclipses or dips in the light curve. Type I X-ray bursts have never been reported for this source, therefore the nature of the compact object is uncertain. The donor star is most probably a CO or ONe white dwarf, based on the presence of oxygen and carbon lines in the optical spectra \citep{nele} and neutral neon in the X-ray spectra of this source with abundance well above the abundance expected in the interstellar medium (ISM) \citep{juett}. \\
The X-ray photons, that are reprocessed in this oxygen-rich material transferred to the accretor may show up in the spectrum in the form of a broad emission O VIII Ly$\alpha$ line (at $\sim0.65$ keV), as recently found using {\it XMM-Newton} data in a similar oxygen-enriched UCXB 4U 0614+091 \citep{madej}. \citet{schulz} confirmed the presence of the OVIII Ly$\alpha$ line in 4U 0614+091 using Chandra data. This line is the second strongest line after Fe K$\alpha$ (at $\sim$6.4 keV) in the reflection spectrum in the case of an accretion disc with solar abundances. However, it may become the most prominent line, when a significant overabundance of oxygen in the accretion disc is considered \citep{ballant}. Since the line is thought to originate from the inner part of the accretion disc it is broadened by the effects of strong gravity close to the accretor, like gravitational redshift and the relativistic Doppler effect \citep{fabian89}. Additional broadening may come from Compton upscattering on hot electrons in the ionized disc material \citep{ballant}.\\
Evidence for an emission feature at $\sim0.7$ keV in 4U 1543-624 was found before in {\it ASCA} and {\it BeppoSAX} data \citep{shulz2003}. \citet{juett} used {\it ASCA} data to show that the feature could be fitted with a higher-than-solar neon abundance in the line of sight. The analysis of the high-resolution data of 4U 1543-624 confirmed the overabundance of neon \citep{juett2003}. The authors reported however a 10-20\% residuals around $\sim0.7$ keV (18~\AA), which they ascribed to an calibration uncertainty. \\
Emission at around 6.4 keV indicating the presence of the Fe K$\alpha$ line was detected for this source in the high state spectra of {\it RXTE} and {\it EXOSAT} \citep{shulz2003}. \citet{juett2003} found also a Gaussian feature in the {\it XMM-Newton} data, discussed in this paper. The parameters of the Gaussian are comparable to the one found by \citet{shulz2003} in {\it RXTE} data. The authors however reject the interpretation of this feature as due to Fe emission, because of the large width and instability of the {\it XMM} fits. Analysis of the same {\it XMM-Newton} data set done by \citet{ng} reveals the presence of a much narrower Fe K$\alpha$ line than that
reported by \citet{juett2003}.\\
We reanalyze the high-resolution and broad band spectra of 4U 1543-624 obtained by the \textit{XMM-Newton} satellite. Since the work of \citet{juett2003}, the calibration of the RGS instrument improved sufficiently to show that the residual at $\sim 0.7$ keV is not a calibration issue, but a real emission feature. Furthermore, the feature is not present in the sources of similar flux and spectrum. In the spectrum obtained with the HETGS instrument on board of \textit{Chandra} satellite we find a residual at $\sim 0.7$ keV similar to the one in the RGS spectrum. We attribute this feature to a broad O VIII Ly$\alpha$ line. Additionally we investigate the presence of the emission feature at $\sim 6.4$ keV in both data sets.
\vspace{-5mm}
\section{Observations and data reduction}
\textit{XMM-Newton} observed 4U 1543-624 for $\sim$ 50 ksec on 2001 February 4 starting at 13:17 UT. During the observation the two Reflection Grating Spectrometers (RGS) and the European Photon Imaging Cameras MOS1, MOS2 and pn were collecting data. The MOS1 and pn cameras were operated in the Timing mode, while MOS2 was operated in the Full Frame (Imaging) mode.  Because of the larger uncertainties in the spectral calibration for the Timing mode with respect to the Imaging mode of the MOS cameras we do not analyze data obtained from MOS1 camera. In the case of pn data, it was reported that pn camera in the Timing mode is not well calibrated in the soft part of the spectrum. Therefore in this analysis the part of the pn spectrum below 1.5 keV is excluded.
We reduce the data using the \textit{XMM-Newton} Data Analysis software SAS version 10.0. 
The observation is not contaminated by soft proton flares.
We extract the MOS2 observation with pixel pattern $\leq$ 12. The MOS2 observation suffers from pile-up, therefore we exclude events from within a circle with a radius of 15 arcsec centered on the source position. We check using the {\sc epat-plot} tool that there is no evidence for pile-up in the selected annulus. The background spectrum for the MOS2 observation is extracted from CCD-3. The net source count rate is $9.37\pm0.01$ c/s. 
We extract the pn observation with pixel pattern $\leq$ 4 in {\sc rawx} from 30 to 45. The background is extracted in {\sc rawx} from 10 to 20. The net source count rate is $92.41\pm0.05$ c/s.
The data collected by RGS are reduced using the standard software pipeline which generates source and background spectra as well as response files. The net source count rate for RGS1 and RGS2 are $4.04\pm0.01$ c/s and $5.75\pm0.01$ c/s, respectively.\\
\textit{Chandra} observed 4U 1543-624 for $\sim$30 ksec on 2000 September 12 starting at 08:16. During the observation the High Energy Transmission Grating Spectrometer (HETGS) together with the Advanced CCD Imaging Spectrometer (ACIS) was used. The HETGS includes two transmission gratings: the Medium Energy Grating (MEG) and High Energy Grating (HEG) covering the wavelength range 2.5-31~\AA\ and 1.2-15~\AA\ respectively. We use processed data available on the Transmission Grating Catalog and Archive (TGCat, http://tgcat.mit.edu/). The net source count rate for MEG is $10.8\pm0.2$ c/s and for HEG is $5.25\pm0.02$ c/s.\\
We fit the data using the {\sc xspec} package \citep{arnaud}. Errors on the fit parameters reported throughout this paper correspond to a 68\% confidence level for each single fit parameter ($\Delta\chi^2=1$). In the case of the {\it Chandra} observation we use the Cash statistics, since most of the spectral bins in the region where the spectral edges occur (see Sec.~3.2) contain less than 30 counts.
\vspace{-3mm}
\section{Analysis and results}
\subsection{Continuum model}
The model reported in the literature for the {\it XMM-Newton} and {\it Chandra} data sets, analyzed in this paper, consists of an absorbed powerlaw and black body \citep{juett2003}. Another model includes an absorbed black body plus a disk black body \citep{ng} and was used to fit only EPIC-pn data, in the energy range 0.7-10 keV. \citet{shulz2003} used the model comprised of black body and Comptonization components to fit {\it RXTE}, {\it BeppoSAX} and {\it ASCA} data. The resulting high optical depths ($\tau > 10$ in most cases) of the Comptonizing plasma mean that the spectra are close in shape to a cut-off powerlaw. Since the limited energy range (until 10 keV)  of EPIC and HETGS data (until 7 keV) does not allow us to constrain the parameters of Comptonization model, we choose combinations of black body and powerlaw components.\\
We simultaneously fit the RGS, MOS2 and pn data obtaining fits with similar $\chi_{\nu}^2$ value using the following models. Model A: a broken powerlaw ({\sc bknpower} model in XSPEC) together with a disc black body, absorbed by cold circumstellar/interstellar matter ({\sc TBnew}). The {\sc TBnew} model is also used in the Model B and Model C. Model B: an absorbed disc black body ({\sc diskbb} in XSPEC) and a black body ({\sc bbodyrad}). Model C: an absorbed cut-off powerlaw ({\sc cutoffpl}) and a black body. The best-fit parameters are listed in the Table~1. In the case of HETGS data, we obtain the best fit to MEG and HEG data (fitted simultaneously) using an absorbed powerlaw and a black body (model D). We add a multiplicative factor ({\sc constant}) in front of all of the models in order to account for possible cross-calibration uncertainties of the instruments. For the reference abundances of elements in the neutral absorber we use the proto-solar abundances of \citet{lodders}.\\
The RGS data are fitted in the energy range from 7~\AA\ to 30~\AA, MOS2 in the range 0.4 keV to 10 keV, pn in the range 1.5 keV to 10 keV, MEG in the range from 2~\AA\ to 26~\AA\ and HEG in the range 2~\AA\ to 13~\AA. The source flux in the energy range 0.5-10.0 keV for the {\it XMM-Newton} observation is $7.37\pm0.01\times10^{-10}$ erg/cm$^2$/s (calculated using MOS2 data and Model C) and for the {\it Chandra} observation is $10.02\pm0.08\times10^{-10}$ erg/cm$^2$/s (calculated using MEG data).
\begin{figure*}  
\hbox{
\psfig{figure=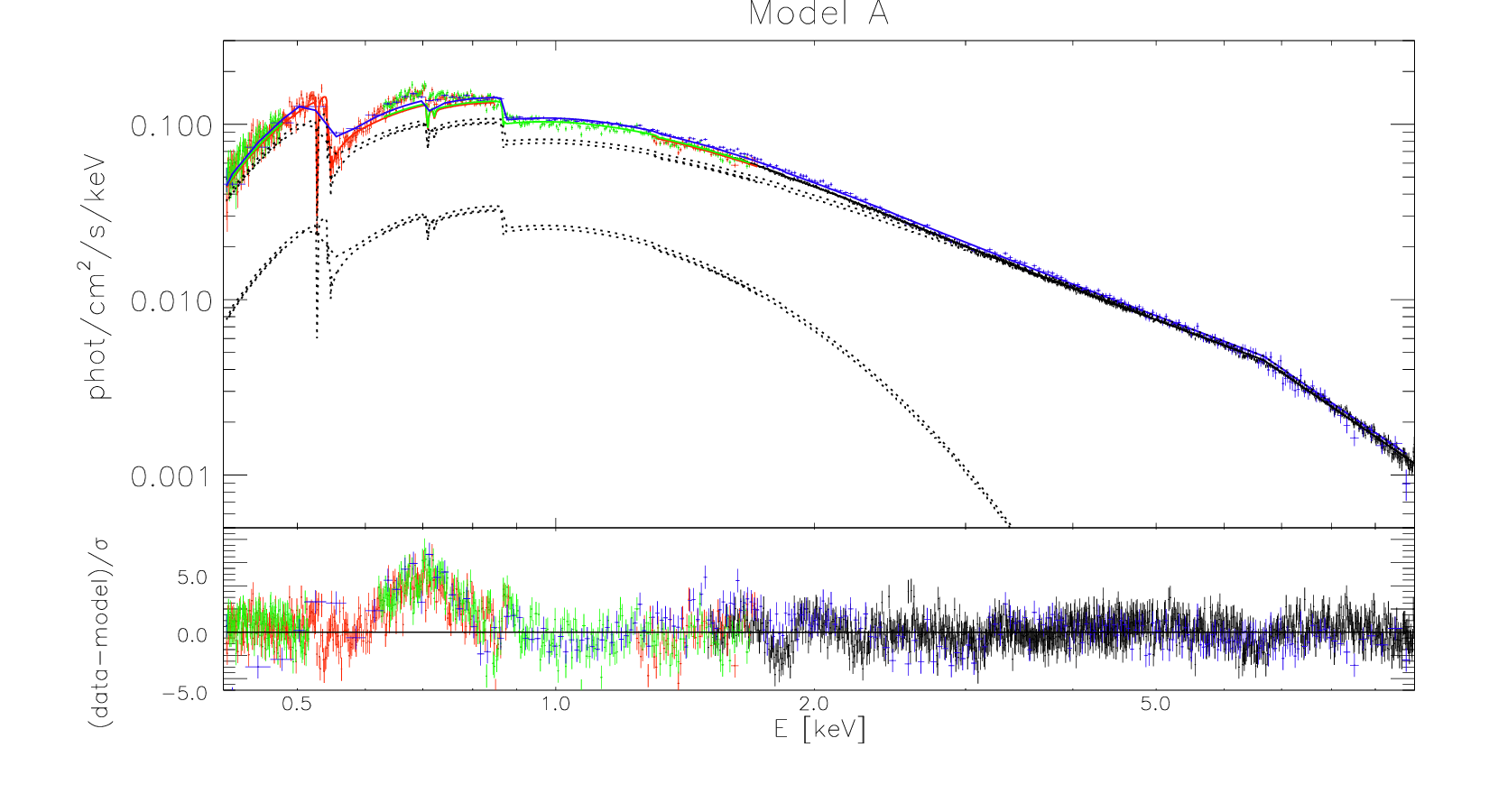,height=5.0cm,width=8.7cm}
\psfig{figure=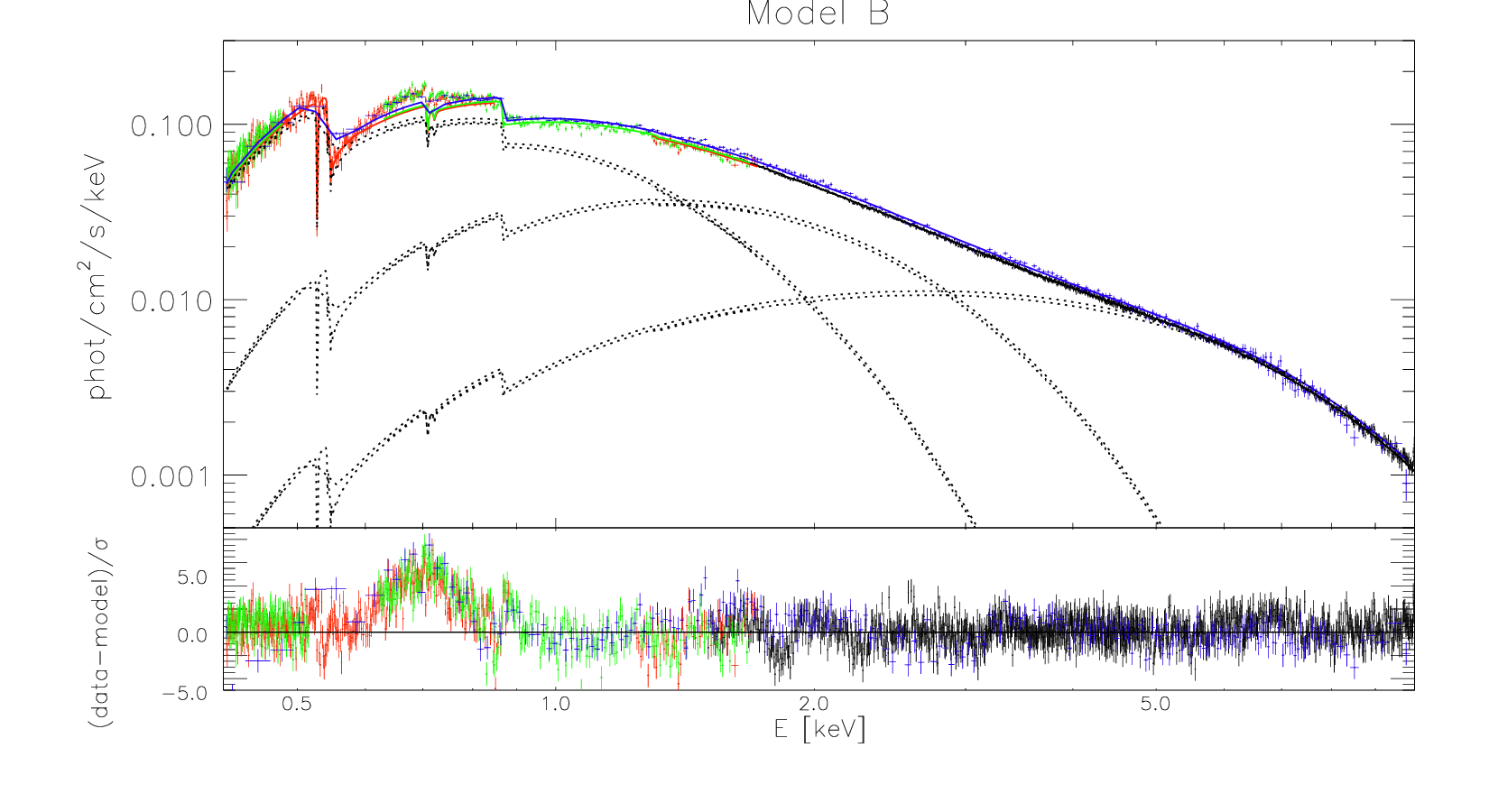,height=5.0cm,width=8.7cm}
}
\hbox{
\psfig{figure=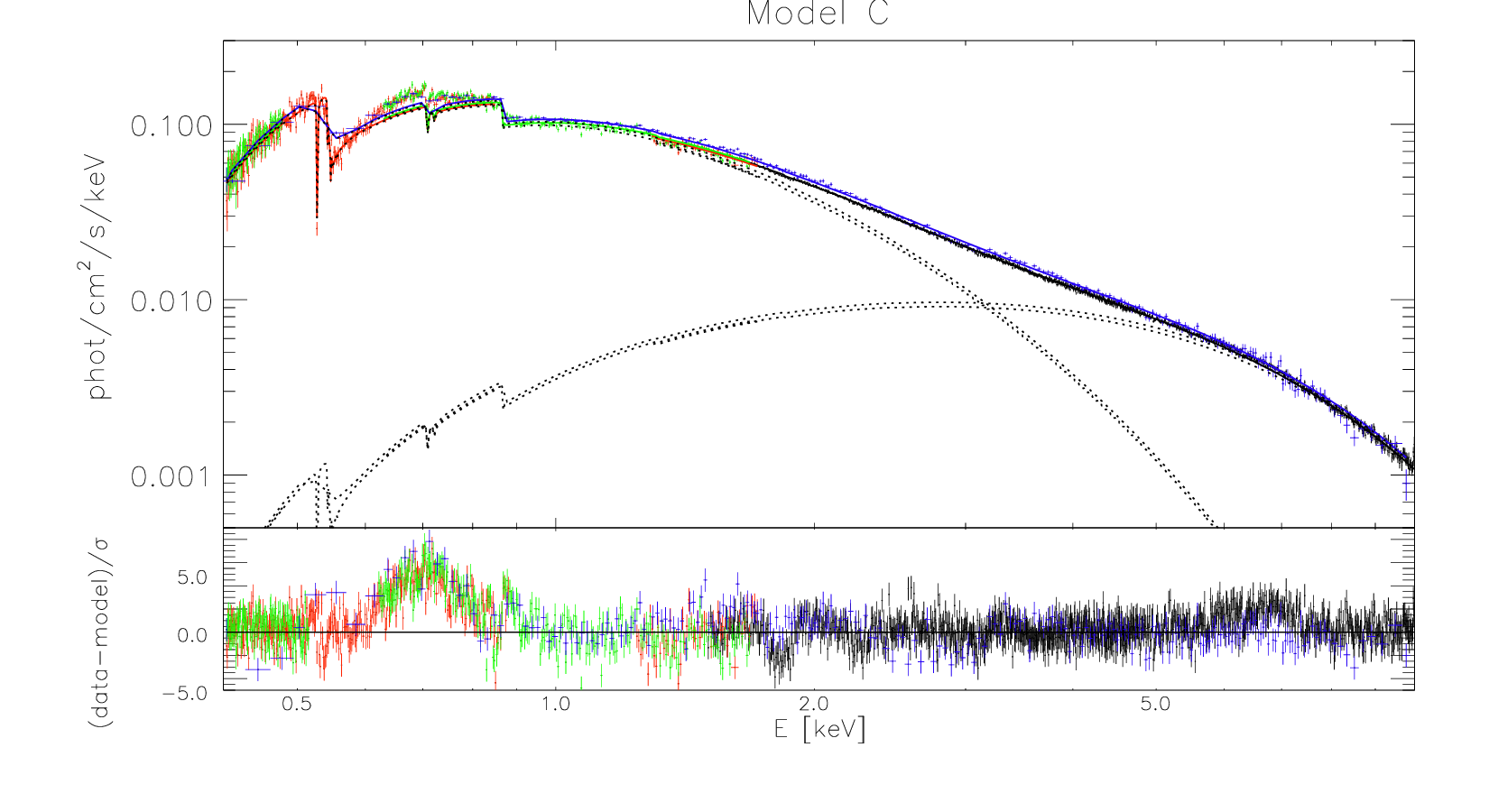,height=5.0cm,width=8.7cm}
\psfig{figure=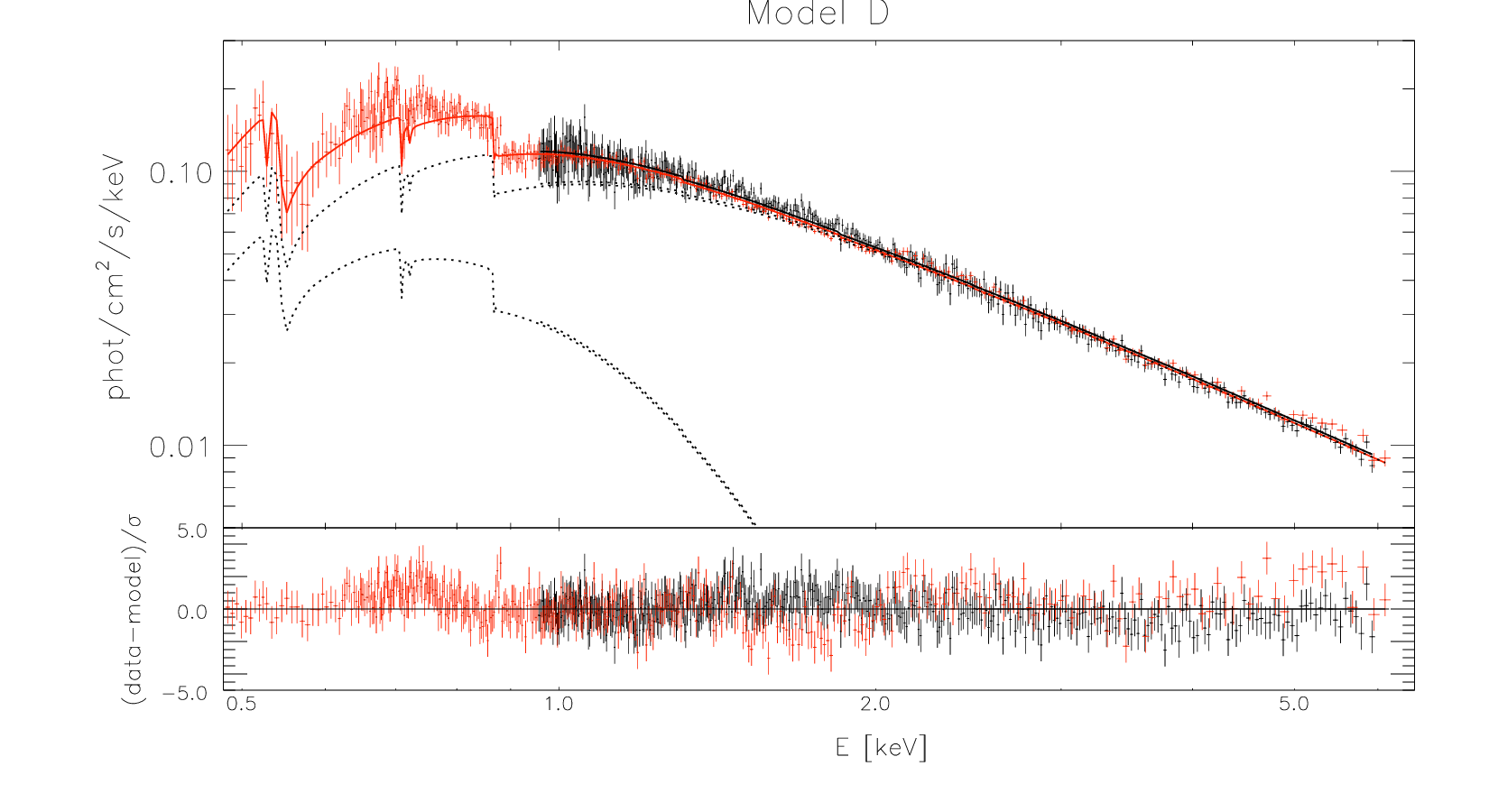,height=5.0cm,width=8.7cm}
}
  \caption{The continuum models fitted to the {\it XMM-Newton}: RGS1 (red) + RGS2 (green)+MOS2 (blue)+pn (black) and {\it Chandra} MEG (red)+HEG (black) spectra of 4U1534-624. The panels show the unfolded spectra and the model components (dotted lines). Top left: {\it XMM-Newton} data set fitted with model A: TBnew*(bknpower+diskbb), Top right: {\it XMM-Newton} data fitted with model B: TBnew*(bbrad+bbrad+diskbb), Bottom left: {\it XMM-Newton} data fitted with model C:  TBnew*(cutoffpl+bbrad). Bottow right: {\it Chandra} data fitted with model D:  TBnew*(pl+bbrad). The residuals show a broad emission feature at $\sim 0.7$ keV and weak feature at $\sim 6.6$ keV when model B and C is used. The absorption features at $\sim1.84$ keV and $\sim2.28$ keV in the pn data are probably caused by calibration uncertainties \citep{ng}. } 

\end{figure*} 
\newcommand{\fivelinebrace}{$\left. \begin{array}{c} \\ \\ \\ \\ \\ \end{array} \right\lbrace$}
\newcommand{\threelinebrace}{$\left. \begin{array}{c} \\ \\ \\ \end{array} \right\rbrace$}
\vspace{-3mm}
\subsection{Measuring abundances in the RGS \& HETGS data}
We start the analysis of the region around O VIII Ly$\alpha$ (at $\sim 18$~\AA$\approx0.7$ keV) from fitting absorption edges prominent in this part of the spectrum (see Fig.~2).
We measure the depth of the following absorption edges: the Ne K-edge (at 14.3~\AA), the Fe L-edge (at 17.3~\AA) and the O K-edge (at 23.05~\AA) by fitting the spectra locally. We use the following wavelength ranges for these fits:  $21-26$~\AA\ for the O K-edge, $16-18$~\AA\ for the Fe L-edge and $12-16$~\AA\ for the Ne K-edge. For these local fits the continuum model is fixed to the one found from the overall fit to the RGS, MOS2 and pn data. \\
In order to determine the depths of the spectral edges of oxygen, iron and neon in MEG and HEG data we use the same wavelength ranges as used for the local fits to the RGS data. We fix the continuum model parameters to those found from the overall fit to the MEG and HEG data. The results are included in the Table~1.
\begin{figure*}  
\hbox{
\psfig{figure=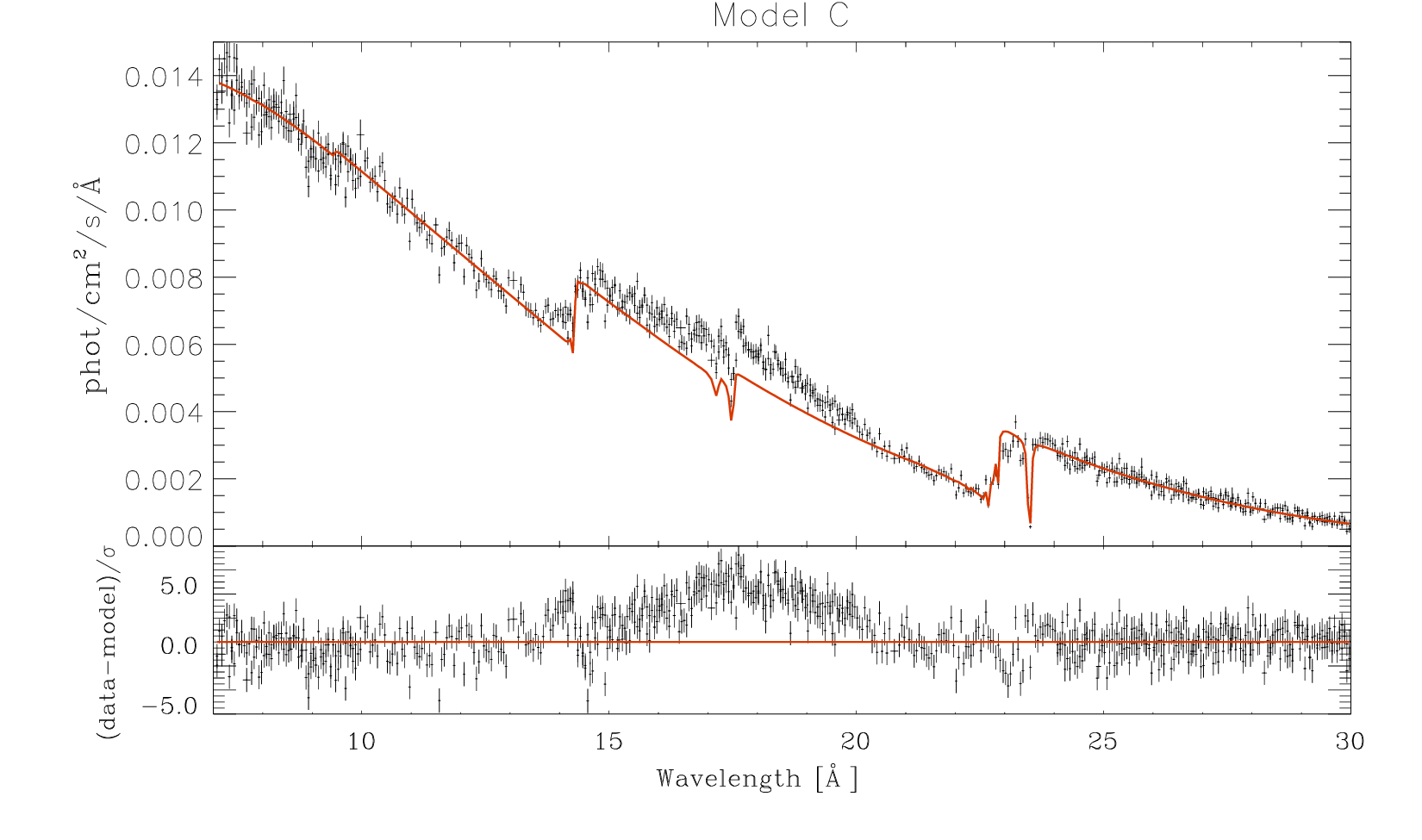,height=5.0cm,width=9.0cm}
\psfig{figure=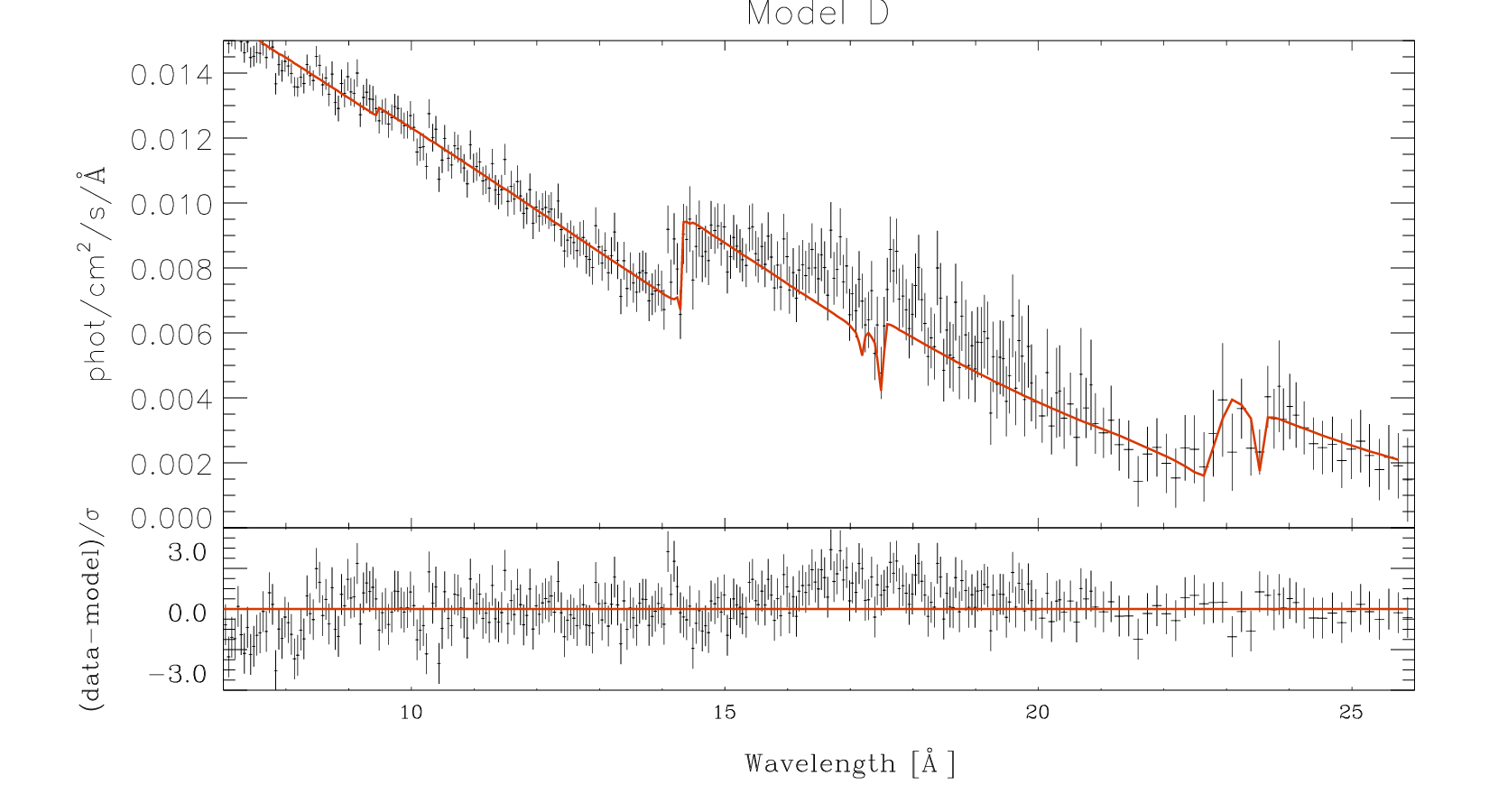,height=5.0cm,width=9.0cm}
}
  \caption{Left: the continuum model C overlaid on the {\sc XMM-Newton} RGS spectrum of the 4U 1543-624. Right: the continuum model D overlaid on the {\sc Chandra} MEG spectrum. Both spectra show prominent absorption edges: O K-edge at 23.05~\AA, Fe L-edge at 17.3~\AA, Ne K-edge at 14.3~\AA. The residuals show a broad emission feature at $\sim0.7$ keV.}
\vspace{-3mm}
\end{figure*} 
\vspace{-3mm}
\subsection{Emission features at $\sim 0.7$ keV and $\sim 6.4$ keV}
We refit the RGS, MOS2 and pn spectrum with the column densities (abundances and $N_{H}$) of oxygen, iron and neon fixed to the values measured from the local fits to the RGS data. We fix the $N_{H}$ to the value obtained from fitting the overall spectrum excluding the region from 16~ \AA\ to 21~\AA. The fits for all investigated continuum models reveal structure around 0.7 keV (see Fig.~1). In the case of model B and C for the continuum there is also an emission feature present around 6.6 keV. We do not detect a significant emission feature at $6.6$ keV while using model A. It is conceivable that this is caused by the fact that the energy of the powerlaw break in this case is close to 6.6 keV. \\
First, we model the emission features with a Gaussian. The centroid of the line for all of the continuum models is around 0.69 keV with $\sigma\approx0.06$ keV. As for the feature in the high energy part of the spectrum, the centroid is around 6.7 keV with $\sigma\approx0.4$ keV and 6.6 keV with $\sigma\approx0.6$ keV for models B and C, respectively. \\
The major improvement of the $\chi_{\nu}^2$ value comes from the inclusion of the line in the soft part of the spectrum ($\Delta\chi^2=-1555,-1854$ for model B and C, respectively, see Table~2). Further, minor improvement of the $\chi_{\nu}^2$ value  comes from inclusion of the line around 6.6 keV ($\Delta\chi^2=-49,-150$ for model B and C, respectively).\\
We fit MEG and HEG data with the abundances of oxygen, iron and neon fixed to the values obtained from the local fits and $N_{H}$ obtained from fitting the overall spectrum excluding the region from 16~\AA\ to 21~\AA. The fit reveals the presence of an emission feature around 18~\AA\ as well. Inclusion of the Gaussian line improves the fit to MEG and HEG data by $\Delta$C$_{stat}=-101$. The center of the Gaussian line is at $\approx0.7$ keV and the $\sigma\approx0.06$ keV. We do not detect a significant emission feature around 6.6 keV.

\begin{table}
\begin{center}
\caption{Parameters from the best fit continuum models to {\it XMM-Newton} (Model A, B, C) and {\it Chandra} data (Model D). Model {\sc A: TBnew*(bknpower+diskbb)}, {\sc B: TBnew*(bbrad+bbrad+diskbb)}, {\sc C: TBnew*(cutoffpl+bbrad)}, {\sc D: TBnew*(powerlaw+bbrad)} The errors on the abundances of oxygen, iron and neon are calculated from the local fits. In this fit the region from 16~\AA\ to 21~\AA\ was excluded.}
\begin{tabular}{lclc}
Parameter & Model A & Parameter & Model B\\ 
\hline
$N_{H}$ $[10^{21} {\rm cm}^{-2}]$&$2.93\pm0.02$ & $N_{H}$ $[10^{21} {\rm cm}^{-2}]$&$2.54\pm0.03$\\
$A_{{\rm O I}}$&$1.11\pm0.02$ & $A_{{\rm O I}}$&$1.31\pm0.03$\\
$A_{{\rm Ne I} }$&$4.4\pm0.1$ & $A_{{\rm Ne I} }$&$5.3\pm0.1$ \\
$A_{{\rm Fe I}}$&$0.82\pm0.09$ & $A_{{\rm Fe I}}$&$1.0\pm0.1$\\
$\Gamma_{1}$ &$1.861\pm0.007$ & $T_{bb1}$ [keV]&$0.515\pm0.004$\\
$\Gamma_{2}$ &$3.36\pm0.03$ & $T_{bb2}$ [keV]&$1.524\pm0.004$\\
$E_{0} $ [keV]&$6.69\pm0.03$ & $T_{{\sc dbb}}$ [keV]&$0.320\pm0.004$\\
$T_{{\sc dbb}}$ [keV]&$0.441\pm 0.005$ & $\chi_{\nu}^{2}\slash d.o.f.$ &1.41/2762\\
$\chi_{\nu}^{2}\slash d.o.f.$ &1.42/2761 & \\
\hline
Parameter & Model C & Parameter & Model D\\ 
\hline
$N_{H}$ $[10^{21} {\rm cm}^{-2}]$&$2.62\pm0.03$ & $N_{H}$ $[10^{21} {\rm cm}^{-2}]$& $3.37\pm0.1$\\
$A_{{\rm O I}}$&$1.25\pm0.03$ & $A_{{\rm O I}}$&$1.0\pm0.2$\\
$A_{{\rm Ne I} }$&$5.1\pm0.1$ & $A_{{\rm Ne I} }$&$4.3\pm0.3$ \\
$A_{{\rm Fe I}}$&$1.0\pm0.1$ & $A_{{\rm Fe I}}$&$0.9\pm0.3$\\
$\Gamma$ &$0.99\pm0.03$ & $\Gamma$ & $1.71\pm0.01$\\
$E_{cut}$ [keV]&$1.14\pm0.02$ & $T_{{\sc bb}}$ [keV]& $0.221\pm0.008$\\
$T_{{\sc bb}}$ [keV]& $1.584\pm0.005$& C$_{stat}\slash d.o.f.\slash N_{BIN}$ &1110/815/821\\
$\chi_{\nu}^{2}\slash d.o.f.$ &1.46/2762 & \\
\end{tabular}
\end{center}
\vspace{-3mm}
\end{table}

\begin{table*}
\begin{center}
\caption{Parameters of the Laor profiles fitted to emission features found at $\sim 0.7$ keV and $6.6$ keV in the {\it XMM-Newton} and {\it Chandra} data. In the case of {\it XMM-Newton} data three different continuum models are used (see Table~1 and Sec. 3.1). The values of abundances are fixed to the values found form local fit and the value of $N_{H}$ is fixed to the one found from the fit with excluded region from 16~\AA\ to 21~\AA. The energies of the lines $E$ and standard deviations $\sigma$ are in keV, equivalent widths in eV. }
\begin{tabular}[t]{lcccccccccc}
\multicolumn{10}{c}{Gaussian}  \\
\hline
Model & $E_{1}$ & $\sigma_{1}$ & $EqW_{1}$ & $\chi^2_{\nu}\slash d.o.f.$ & $\Delta\chi$ & $E_{2}$ & $\sigma_{2}$  & $EqW_{2}$ & $\chi^2_{\nu}\slash d.o.f.$ & $\Delta\chi$\\
\hline
A &-&-&-&1.89/2920&-&-&-&-&-&- \\

&$0.691\pm0.002$&$0.055\pm0.002$&$27.2\pm0.9$&1.42/2917&-1415&-&-&-&-&-\\

\hline
B &-&-&-&1.93/2920&-&-&-& -&-&-\\
 &$0.689\pm0.002$&$0.058\pm0.002$&$31\pm1$&1.40/2917&-1555&$6.73\pm0.06$&$0.37\pm0.05$&$26\pm11$&1.38/2914&-49\\

\hline
C &-&-&-& 2.05/2921&-&-&-&-&-\\
 &$0.691\pm0.002$&$0.064\pm0.002$&$35\pm1$&1.41/2918&-1854&$6.55\pm0.05$&$0.61\pm0.05$&$67\pm19$&1.36/2915&-150\\
\hline
& &  & & $C_{\rm stat}\slash d.o.f.\slash N_{BIN}$ & $\Delta$C &  &   & &  & \\
D &-&-&-&1316/913/918&-&-&-&-&-&-\\
&$0.70\pm0.01$&$0.057\pm0.008$&$30\pm3$&1215/910/918&-101&-&-&-&-&-\\
\end{tabular}
\end{center}
\vspace{-3mm}
\end{table*}

\begin{table}
\begin{center}
\caption{Parameters of the Laor profiles fitted to emission features found at $\sim 0.7$ keV and $6.6$ keV in the {\it XMM-Newton} and {\it Chandra} data. In the case of {\it XMM-Newton} data three different continuum models are used (see text). The energy of the line marked in italics is fixed during the fitting.}
\begin{tabular}[t]{lcccc}
\multicolumn{5}{c}{Laor} \\
\hline
Model & $E$ [keV] & $i$ [deg] & $r_{in}$ [$GM/c^2$]& $\chi^2_{\nu}\slash d.o.f.$\\
\hline
A&{\it 0.654}&$70\pm1.5$&$13.0^{+0.4}_{-0.8}$& 1.46/2916\\
\hline
B &{\it 0.654}&$86.9\pm0.2$&$4.1\pm0.4$&1.44/2916\\
&$6.3\pm0.1$&$87.0\pm0.2$&$4.4\pm0.4$&1.43/2914$^{*}$\\
\hline
C &{\it 0.654}&$86.7\pm0.1$&$3.8\pm0.2$&1.45/2917\\
&$6.38\pm0.09$&$86.8\pm0.1$&$3.6\pm0.3$&1.42/2915$^{*}$\\
\hline
&&&& $C_{\rm stat}\slash d.o.f.\slash N_{BIN}$\\
D&{\it 0.654}&$54\pm1$&$3.1\pm0.7$&1216/909/918\\
\end{tabular}
\end{center}
{\footnotesize $^{*}$ In this fit the Laor profile used for modelling the feature at $\sim 0.7$ keV is also included. }
\vspace{-3mm}
\end{table}
\begin{table}
\begin{center}
\caption{Parameters from the best-fit model consisting of relativistically broadened reflection component, cut-off powerlaw and black body, fitted to the RGS1,2+MOS2+pn data. Parameters fixed during the fitting are marked in italics.}
\begin{tabular}{lc}
\multicolumn{2}{c}{\sc TBnew*(kdblur*reflionx+bbody+cutoffpl)} \\
\hline
$N_{H}$ $[10^{21} {\rm cm}^{-2}]$&$2.64\pm0.04$\\
$A_{{\rm O I}}$&$1.26\pm0.02$\\
$A_{{\rm Ne I} }$& $5.4\pm0.1$\\
$A_{{\rm Fe I}}$&$0.66\pm0.09$\\
$\Gamma$ &$0.50\pm0.05$\\
$E_{cut}$ [keV]&$0.9\pm0.1$\\
$T_{{\sc bb}}$ [keV]&$ 1.481\pm0.007$\\
$q_{\sc kdblur}$ & $2.77\pm0.03$\\
$r_{in} $ [$GM/c^2$]& {\it 6.0}\\
$i$ [deg]&$54.3\pm0.4$\\
$\Gamma_{\sc reflionx}$ & $1.97\pm0.04$\\
$A_{Fe}^{*}$ & {\it 0.5} \\
$\xi$ [erg cm/s]&$250\pm11$\\
$\chi_{\nu}^{2}\slash d.o.f.$ &1.44/2913\\
\end{tabular}
\end{center}
{\footnotesize $^{*}$ Abundance of iron in the {\sc refionx} model. Abundances of other elements in this model are fixed at solar values.}
\vspace{-3mm}
\end{table}

  \begin{figure}  
\vspace{-3mm}
\psfig{figure=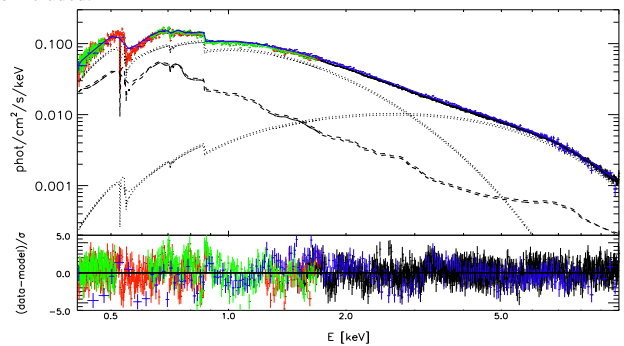,height=5.0cm,width=9.0cm}
  \caption{The best-fitting continuum model consisting of an absorbed reflection component (dashed line), cut-off powerlaw and black body (dotted lines) overlaid on the RGS and pn spectrum of the 4U 1543-624. The upper frame shows the spectrum and model components, the lower frame shows residuals in the units of the standard deviation. } 
\vspace{-3mm}
\end{figure} 
\subsection{Laor profile}
Based on the fact that the accretion disc is overabundant with oxygen and affected by the strong gravitational field of the accretor, a possible interpretation of the feature at $\sim0.7$ keV is a relativistically broadened emission line of O VIII Ly$\alpha$. Therefore we fit a Laor profile to the residuals present in the RGS+MOS2 and HETGS data. We fix the wavelength of the line to 0.654 keV (the position of O VIII Ly$\alpha$). The outer radius of the disc is fixed to 1000 $GM/c^2$. The fit to RGS+MOS2 data gives an inclination of $\sim87^{\circ}$ and an inner radius of $\sim4\ GM/c^2$ for model B and C and $\sim 13 GM/c^2$ for model A. \\
Fitting the residual present in the HETGS data with the Laor profile gives a reasonable value of the inclination of $54^{\circ}$ (see Table 3). \\
Additionally we fit the Laor profile to both features at $\sim0.7$ and $\sim6.6$ keV for model B and C, coupling the inner radius, inclination and emissivity index. The energy of the line in the soft part of the spectrum is fixed to the energy of OVIII Ly$\alpha$. The energy of the line at $\sim6.6$ keV we keep free due to the possible contribution from a number of iron ions (Fe I-Fe XXVI). In the case of both models B and C we obtain a result consistent with the case of the fit of a single Laor profile to the 0.7 keV feature for the inclination and inner radius. \\
Similar to the case of 4U 0614+091, the lines, apart from relativistic broadening, can be additionally affected by the Compton scattering in the ionized disc material \citep{madej}, which, if not taken into account, could artificially increase the inclination in the Laor profile fit. 
\vspace{-5mm}
\subsection{Reflection model}
We use the RGS, MOS2 and pn data to fit a relativistically broadened reflection model {\sc reflionx}, available in {\sc xspec}, which takes into account effects of Compton scattering in the ionized reflecting material. Relativistic broadening is modeled using {\sc kdblur} and for the Galactic absorption we choose again the {\sc tbnew} model \citep{wilms}. For the continuum model we choose the cut-off powerlaw and the disc black body ({\sc TBnew*(kdblur*reflionx+bbody+cutoffpl)}, see Fig.~3). The only adjustable abundance in the {\sc reflionx} model is the abundance of iron. Therefore, in order to mimic the enhancement of oxygen we reduce the abundance of iron by a factor of two. During the fit abundances of interstellar oxygen, neon, iron and $N_{H}$ are kept free. We obtain a fit with $\chi_{\nu}^2=1.44$ for 2913 d.o.f. (see Table~4). The inclination is $\sim 54^{\circ}$ and the ionization parameter is $\sim 250$, which indicates that the line is broadened by Compton scattering as well. 
\vspace{-5mm}
\section{Discussion}
We report the discovery of a broad emission feature near 0.7 keV in the high-resolution RGS, broad band MOS2 and HETGS data of 4U 1543-624. A similar feature was found recently in the oxygen enriched UCXB 4U 0614+091 \citep{madej,schulz}. A reasonable interpretation of this feature is that of a relativistically broadened O VIII Ly$\alpha$ line, which would be the most prominent line in the reflection spectrum in the case of an oxygen-rich accretion disc. Fitting a Laor profile to {\it XMM} data gives a high inclination, inconsistent with the lack of eclipses and dips in the light curve of the source. The line is most probably also affected by Compton scattering in the ionized material of the disc. Fitting a relativistically broadened reflection model we obtain a more reasonable value of the inclination of $\sim 54^{\circ}$ and significant Compton broadening. \\
We have checked the spectra for the presence of an emission feature at $\sim 6.6$ keV and we have found a Gaussian shaped feature, when the continuum is described by the model that includes a disk black body and two black body emitters or black body and cut-off powerlaw. The centroid of the feature is at $\sim 6.6$ keV with $\sigma\approx0.5$ keV, which is consistent with the values obtained by \citet{ng}. The energy centroid indicates emission from iron ions up to Fe XXV. However, the value of the energy found from fitting a Laor profile is $\sim 6.4$ keV, which suggests emission from iron ions from Fe I to Fe XIX. We do not detect significant emission at the position of the Fe K$\alpha$ line when disk black body plus broken powerlaw fit function is used. The energy of the powerlaw break however falls in the range of the iron K line emission, potentially masking its presence. We do not detect any significant emission feature at $\sim 6.6$ keV in HETGS data. The same conclusion was reached by \citet{juett2003}.\\ 
Concerning the ISM/CSM absorption in the line of sight, we find an overabundance of neon in the absorber with respect to \citet{lodders} in RGS and HETGS data, which corroborates the previous measurements \citep{juett,juett2003}. A residual around the oxygen edge suggests that oxygen may be also in molecular form, for which the fine structure of the absorption edge can be significantly different compared to the atomic oxygen \citep{pinto}. The residual near the neon edge may indicate the presence of warm ionized gas in the line of sight  \citep[similarly to the case of Cyg X-1,][]{juett2006}. 
\vspace{-5mm}
\section*{Acknowledgments}
We thank Frank Verbunt and the referee for useful comments. PGJ acknowledges support from a VIDI grant from the Netherlands Organisation for Scientific Research.
\vspace{-5mm}
 \bibliographystyle{mn2e}
 \bibliography{mybib15}

%\begin{thebibliography}{99}

%\end{thebibliography}

\end{document}